\documentclass[11pt]{article}
\begin{document}
\title{Non-entropic theory of rubber elasticity:\\
flexible chains with weak excluded-volume interactions}

\author{A.D. Drozdov\footnote{
Phone: 972-86472146.
E-mail: aleksey@bgumail.bgu.ac.il}\\
Department of Chemical Engineering\\
Ben-Gurion University of the Negev\\
P.O. Box 653\\
Beer-Sheva 84105, Israel}
\date{}
\maketitle

\begin{abstract}
Strain energy density is calculated for a network of
flexible chains with weak excluded-volume interactions
(whose energy is small compared with thermal energy).
Constitutive equations are developed for an incompressible
network of chains with segment interactions at finite
deformations.
These relations are applied to the study of uniaxial
and equi-biaxial tension (compression), where the
stress--strain diagrams are analyzed numerically.
It is demonstrated that intra-chain interactions
(i) cause an increase in the Young's modulus of the
network and
(ii) induce the growth of stresses (compared to an
appropriate network of Gaussian chains), which becomes
substantial at relatively large elongation ratios.
The effect of excluded-volume interactions on the
elastic response strongly depends on the deformation mode,
in particular, it is more pronounced at equi-biaxial tension
than at uniaxial elongation.
\end{abstract}
\vspace*{5 mm}

\noindent
{\bf Key-words:}
Flexible chain,
Excluded-volume interaction,
Polymer network,
Finite deformation,
Path integral
\vspace*{5 mm}

\section{Introduction}

This study is concerned with the elastic response of
polymer networks at finite strains.
According to the classical theory of rubber elasticity \cite{Tre75},
an arbitrary chain in a network is treated as Gaussian,
which allows a simple formula to be derived for the strain
energy density of a network, and stress--strain relations
to be developed in the analytical form.
Two shortcomings of the concept of Gaussian chains
are traditionally emphasized:
(i) this model does not account for long-range interactions
between segments (Gaussian chains can intersect themselves),
and (ii) the end-to-end distance of a Gaussian chain
exceeds its contour length with a non-zero probability.
To avoid these disadvantages, it seems enticing
to replace Gaussian chains in a network by flexible chains
with excluded-volume interactions (this model does not
permit self-intersections) or by semi-flexible chains
(this approach guarantees that the end-to-end distance of a
chain is always less than its contour length).
Although the necessity to go beyond Gaussian chains
has been realized for a long time (the seminal paper by
Flory \cite{Flo49} appeared more than half a century
ago), our knowledge of the mechanical behavior of networks
of non-Gaussian chains remains rather limited, due to
some difficulties in their mathematical treatment.
Serious progress in the analysis of statistics
of single polymer chains and membranes with excluded-volume
interactions was reached by using the renormalization
group technique \cite{CO02,BI03,CO04,Dup04}.
These methods, however, have been employed for the analysis
of the distribution functions only and have not yet been
applied to determine the strain energy of a network,
despite the importance of the latter problem for applications,
see \cite{CJ90,Sch99} and the references therein,
as well as recent publications
\cite{BW03,CG04,KM04,LLH04}.

This study is motivated by a problem which, at first
glance, appears to be quite simple.
Consider an incompressible permanent network
of Gaussian chains with a given set of parameters
(segment length $b_{0}$,
contour length $L$,
number of chains per unit volume $M$)
at a fixed absolute temperature $T$.
Suppose that Gaussian chains in the network
are replaced by flexible chains (with the same
parameters $b_{0}$, L and $M$) with excluded-volume
interactions (whose strength $v_{0}$ will be
defined later).
The question is how the presence of intra-chain interactions
affects the Young's modulus $E$ of the network?
In a more general context, this question may be
reformulated as what is the influence of segment
interactions on the stress--strain relations for
the network?

In order to shed some light on this issue,
it is necessary to define unambiguously what flexible
chains with excluded-volume interactions mean.
Two approaches are conventionally employed to describe
configurations of a polymer chain.
According to the first, a chain is thought of as a random
walk with a small step length $b_{0}$ and a large number
of steps $N$ (their product $L=Nb_{0}$ is assumed to
be finite when $N\to \infty$).
Excluded-volume interactions between segments are
treated as a constraint that rules out
trajectories that intersect themselves (self-avoiding
random walks),
which implies that the ``strength" of segment interactions
$v_{0}$ has no physical meaning.
A disadvantage of this concept is that the distribution
function for end-to-end vectors ${\bf Q}$ of self-avoiding
chains is unknown.
Some approximations for this function are available
at ${\bf Q}\to \infty$ \cite{CJ90}, but they are
insufficient for the determination of elastic moduli.
Nevertheless, the effect of intra-chain interactions
on the elastic moduli may be assessed qualitatively:
as the number of configurations for a self-avoiding
walk is lower than that for a walk without constraints,
and the free energy of a chain is proportional to
its entropy (which, in turn, is proportional to
the logarithm of the number of available configurations),
it is plausible to assume that excluded-volume interactions
between segments reduce the stiffness.

According to the other approach, a chain is treated
as a curve with length $L$ in a three-dimensional space.
Any configuration of the chain is described by the
equation ${\bf r}={\bf r}(s)$, where ${\bf r}$ stands
for the radius vector, and $s\in [0,L]$.
This configuration is characterized by a weight (energy),
which is determined by some Hamiltonian $H({\bf r})$.
For a Gaussian chain, the Hamiltonian reads
\begin{equation}
H_{0}({\bf r})=\frac{3k_{\rm B}T}{2b_{0}}\int_{0}^{L}
\Bigl (\frac{d{\bf r}}{ds}(s)\Bigr )^{2} ds,
\end{equation}
while for a chain with excluded-volume interactions,
this functional is given by \cite{DE86}
\begin{equation}
H({\bf r})=H_{0}({\bf r})+\Phi({\bf r})
\end{equation}
with
\begin{equation}
\Phi({\bf r})=\frac{v_{0}}{2L^{2}}
\int_{0}^{L}\int_{0}^{L} \delta
\Bigl ({\bf r}(s)-{\bf r}(s^{\prime})\Bigr )
ds ds^{\prime}.
\end{equation}
Here $k_{B}$ is Boltzmann's constant,
$\delta({\bf r})$ denotes the Dirac delta-function,
and the pre-factor $v_{0}$ characterizes strength of
segment interactions.

Unlike the concept of random walks, an assessment of
the influence of segment interactions on the Young's
modulus $E$ becomes non-trivial in this case.
On the one hand, the presence of the second term on
the right-hand side on Eq. (2) reduces the number of
configurations with a noticeable weight, which results
in a decrease in the free energy.
On the other hand, this term increases the energy $H$ of
any available configuration, which causes the growth
of the average energy of a chain, and, as a consequence,
an increase in the strain energy density of a network.

A rigorous treatment of the interplay between these two
factors is the objective of this paper.
Due to some technical difficulties in the evaluation
of path integrals for the Hamiltonians (2) and (3),
we confine ourselves to the analysis of weak
excluded-volume interactions, whose energy $\Phi({\bf r})$
is small compared with thermal energy $k_{B}T$.

The exposition is organized as follows.
Section 2 deals with the free energy of a flexible
chain and the strain energy density of a permanent
network of polymer chains calculated within
the entropic and non-entropic concepts.
The free energy of a flexible chain with a small,
but arbitrary functional $\Phi({\bf r})$ is found in Section
3 in terms of an appropriate correlation function.
Section 4 has a merely technical character.
We derive some explicit expressions for correlation
functions, which are employed in Section 5
to determine the free energy of a chain with excluded-volume
interactions.
Constitutive equations for a network of chains with
weak segment interactions are developed in Section 6,
where they are applied to the analysis of uniaxial
and equi-biaxial tension of an incompressible medium.
Some concluding remarks are formulated in Section 7.

\section{The concept of non-entropic elasticity}

We begin with a brief exposition of the classical
theory of rubber elasticity \cite{Tre75},
demonstrate its shortcomings, and introduce some
refinement of the conventional approach.
The concept of entropic elasticity is grounded on the
treatment of a polymer chain as a random walk in a
three-dimensional space.
For definiteness, we suppose that the walk begins at the
origin and has a fixed length $b_{0}$ of each step.
The main hypothesis of this theory is that
the distribution function $p({\bf Q})$
of end-to-end vectors ${\bf Q}$ entirely describes
configurations of a chain.
The free energy of a chain $\Psi({\bf Q})$ is
connected with the distribution function $p({\bf Q})$
by the Boltzmann equation
\begin{equation}
p({\bf Q})=\exp \Bigl (-\frac{\Psi({\bf Q})}{k_{B}T}\Bigr ).
\end{equation}
In the nonlinear elasticity theory, two states of a
medium are distinguished:
(i) the reference (initial) state occupied before
application of external loads,
and (ii) the actual (deformed) state that is acquired
after deformation.
As a polymer chain is entirely characterized by the
relative positions of its end-points, two vectors
are introduced:
the end-to-end vector in the reference state ${\bf Q}$,
and that in the actual state ${\bf Q}^{\prime}$.
These quantities obey the equality
\begin{equation}
{\bf Q}^{\prime}={\bf F}\cdot {\bf Q},
\end{equation}
where ${\bf F}$ is a deformation gradient,
and the dot stands for inner product.
It follows from Eqs. (4) and (5) that the increment of
free energy
\[
\Delta \Psi({\bf F},{\bf Q})
=\Psi({\bf Q}^{\prime})-\Psi({\bf Q})
\]
driven by deformation of the chain reads
\[
\Delta \Psi ({\bf F},{\bf Q})= k_{B}T \Bigl [
\ln p ({\bf Q}) -\ln p({\bf F}\cdot {\bf Q})\Bigr ].
\]
The strain energy per chain $W({\bf F})$ is
determined by averaging the increment of free energy
over the initial distribution of end-to-end vectors,
\begin{equation}
W({\bf F})=k_{B}T \int \Bigl [
\ln p ({\bf Q})-\ln p({\bf F}\cdot {\bf Q})
\Bigr ]p({\bf Q}) d{\bf Q},
\end{equation}
where the integration is performed over the entire space.
Given a strain energy $W$, stress--strain relations for
a chain are determined by conventional formulas,
see, e.g., \cite{Dro96}.
Equation (6) is noticeably simplified when the distribution
function $p({\bf Q})$ is isotropic:
\[
p({\bf Q})=p_{\ast}(Q),
\qquad
Q=|{\bf Q}|.
\]
Bearing in mind that
$|{\bf F}\cdot {\bf Q}|
=({\bf Q}\cdot {\bf C}\cdot {\bf Q})^{\frac{1}{2}}$,
where
\[
{\bf C}= {\bf F}^{\top}\cdot {\bf F}
\]
is the right Cauchy-Green deformation tensor,
and $\top$ stands for transpose,
and introducing a spherical coordinate frame
$\{ Q,\phi, \theta \}$,
we find that
\begin{equation}
W({\bf C})=k_{B}T \int_{0}^{\infty}
p_{\ast}(Q) Q^{2} dQ
\int_{0}^{2\pi} d\phi
\int_{0}^{\pi} \Bigl [\ln p_{\ast} (Q)
-\ln p_{\ast}\Bigl (({\bf Q}\cdot {\bf C}
\cdot {\bf Q})^{\frac{1}{2}}\Bigr )
\Bigr ] \sin \theta d\theta.
\end{equation}
In particular, for a Gaussian chain with the radial
distribution function
\begin{equation}
p_{\ast}(Q)=\Bigl (\frac{3}{2\pi b^{2}}\Bigr )^{\frac{3}{2}}
\exp \Bigl (-\frac{3Q^{2}}{2b^{2}}\Bigr ),
\end{equation}
where $b=\sqrt{b_{0}L}$ is the mean square end-to-end distance,
Eq. (7) implies the classical formula
\begin{equation}
W({\bf C})=\frac{1}{2} k_{B}T \Bigl ({\cal I}_{1}
({\bf C})-3 \Bigr ),
\end{equation}
where ${\cal I}_{m}$ stands for the $m$th principal invariant
of a tensor.

According to the other way of modeling a polymer chain
\cite{DE86}, each configuration is associated
with a curve ${\bf r}(s)$ in a three-dimensional space.
For a chain that begins at the origin and finishes
at a point ${\bf Q}$, the radius vector ${\bf r}(s)$
satisfies the boundary conditions
\begin{equation}
{\bf r}(0)={\bf 0},
\qquad
{\bf r}(L)={\bf Q}.
\end{equation}
The Green function (propagator) of a chain
whose energy is described by a Hamiltonian $H({\bf r})$
reads
\begin{equation}
G({\bf Q})=\int_{{\bf r}(0)={\bf 0}}^{{\bf r}(L)={\bf Q}}
\exp \biggl (-\frac{H({\bf r}(s))}{k_{\rm B}T}\biggr )
{\cal D} [{\bf r}(s)],
\end{equation}
where the path integral with the measure ${\cal D}[{\bf r}]$
is calculated over all curves ${\bf r}(s)$
that obey Eq. (10).
As the functional integral is determined up to
an arbitrary multiplier \cite{Kle95},
the additional restriction is imposed on the function
$G({\bf Q})$,
\begin{equation}
\int G({\bf Q}) d{\bf Q}=1 ,
\end{equation}
which ensures that the Green function $G({\bf Q})$
coincides with the distribution function of
end-to-end vectors $p({\bf Q})$.

Within the entropic elasticity theory,
the strain energy $W$ of a flexible chain with
a Hamiltonian $H$ is determined by using the same
technique as for a chain treated as a random walk:
given $H({\bf r})$,
the Green function $G({\bf Q})$ is calculated
from Eq. (11) and is normalized with the help of Eq. (12)
to obtain the distribution function $p({\bf Q})$.
Afterwards, the strain energy $W({\bf F})$
is determined by Eq. (6).

Two shortcomings of this approach should be mentioned:
(i) for a chain with a Hamiltonian $H$, a correct measure
of the free energy is the average Hamiltonian,
while the use of Eqs. (6), (11) and (12) appears to be
unnecessary and questionable,
and (ii) our previous analysis of a flexible chain
grafted on a rigid surface demonstrates that
the entropic elasticity theory leads to conclusions that
contradict physical intuition \cite{Dro04}.

In this study, we associate the free energy
$\tilde{\Psi}({\bf Q})$ of a chain with
an end-to-end vector ${\bf Q}$ with the
average Hamiltonian of this chain
\begin{eqnarray}
\tilde{\Psi}({\bf Q}) &=&
\langle H\rangle_{\bf Q}
\nonumber\\
&=& \frac{1}{G({\bf Q})}
\int_{{\bf r}(0)={\bf 0}}^{{\bf r}(L)={\bf Q}}
H({\bf r}(s)) \exp \biggl (-\frac{H({\bf r}(s))}
{k_{\rm B}T}\biggr ) {\cal D} [{\bf r}(s)].
\end{eqnarray}
Given $\tilde{\Psi}({\bf Q})$, we determine the normalized
end-to-end distribution function $\tilde{p}({\bf Q})$
from the equation similar to Eq. (4),
\begin{equation}
\tilde{p}({\bf Q})= \exp
\Bigl (-\frac{\tilde{\Psi}({\bf Q})}{k_{B}T}\Bigr )
\biggl [ \int \exp
\Bigl (-\frac{\tilde{\Psi}({\bf Q})}{k_{B}T}\Bigr )
d{\bf Q}\biggr ]^{-1},
\end{equation}
and calculate the strain energy $\tilde{W}({\bf F})$ as
the average (over the distribution function)
increment of free energy,
\begin{equation}
\tilde{W}({\bf F})=\int \Bigl [
\tilde{\Psi}({\bf F}\cdot {\bf Q})
-\tilde{\Psi}({\bf Q})\Bigr ]
\tilde{p}({\bf Q}) d{\bf Q}.
\end{equation}
Simple algebra reveals that for a Gaussian chain
with Hamiltonian (1),
our approach coincides with the conventional one
and results in the strain energy density (9).
However, for a flexible chain with excluded-volume
interactions, Eqs. (13) to (15) differ from appropriate
relations developed within the entropic elasticity theory.

An important remark regarding Eqs. (13) to (15)
is that an additive constant in the expression
for the free energy $\tilde{\Psi}({\bf Q})$
does not affect the strain energy density $\tilde{W}({\bf F})$
(as it is expected).
This constant is excluded from the formula for the
distribution function $\tilde{p}({\bf Q})$ by
the normalization condition (14), whereas it disappears
in Eq. (15) because only the increment of free energy is
substantial for the determination of $\tilde{W}({\bf F})$.

\section{Free energy of a chain with weak
segment interactions}

Our aim now is to calculate the free energy of a
flexible chain with weak intra-chain interactions,
Eqs. (2) and (3), in the first approximation with
respect to the ratio $v_{0}/k_{B}T$.
Inserting expressions (2) and (3) into Eq. (11),
expanding the exponent into the Taylor series,
and disregarding terms beyond the first order of
smallness, we obtain
\begin{equation}
G({\bf Q}) = \int_{{\bf r}(0)={\bf 0}}^{{\bf r}(L)={\bf Q}}
\Bigl ( 1-\frac{\Phi({\bf r}(s))}{k_{B}T}\Bigr )
\exp \biggl (-\frac{H_{0}({\bf r}(s))}{k_{\rm B}T}\biggr )
{\cal D} [{\bf r}(s)]
= \Bigl (1-\frac{\langle \Phi\rangle_{\bf Q}^{0}}{k_{B}T}\Bigr )
G_{0}({\bf Q}),
\end{equation}
where $G_{0}({\bf Q})$ is the Green function for a Gaussian
chain, and the superscript index zero stands for
averaging with the help of the Gaussian Hamiltonian.
It follows from Eqs. (2), (3) and (13) that
\[
\langle H\rangle_{\bf Q} = \frac{1}{G({\bf Q})}
\int_{{\bf r}(0)={\bf 0}}^{{\bf r}(L)={\bf Q}}
\Bigl ( H_{0}({\bf r}(s))+\Phi({\bf r}(s))\Bigr )
\exp \biggl [-\frac{1}{k_{\rm B}T}
\Bigl ( H_{0}({\bf r}(s))+\Phi({\bf r}(s))\Bigr )
\biggr ] {\cal D} [{\bf r}(s)].
\]
Expanding the exponent into the Taylor series,
using Eq. (16), and neglecting terms beyond the first
order of smallness, we find that
\begin{eqnarray}
\langle H\rangle_{\bf Q} &=&
\Bigl (1+\frac{\langle \Phi\rangle_{\bf Q}^{0}}{k_{B}T}\Bigr )
\frac{1}{G_{0}({\bf Q})}
\int_{{\bf r}(0)={\bf 0}}^{{\bf r}(L)={\bf Q}}
\Bigl ( H_{0}({\bf r}(s))+\Phi({\bf r}(s))-\frac{1}{k_{B}T}
H_{0}({\bf r}(s))\Phi({\bf r}(s)) \Bigr )
\nonumber\\
&&\times \exp \biggl [-\frac{1}{k_{\rm B}T}
H_{0}({\bf r}(s))\biggr ] {\cal D} [{\bf r}(s)]
\nonumber\\
&=&\Bigl (1+\frac{\langle \Phi\rangle_{\bf Q}^{0}}{k_{B}T}\Bigr )
\Bigl (\langle H_{0}\rangle_{\bf Q}^{0}
+\langle \Phi\rangle_{\bf Q}^{0}
-\frac{1}{k_{B}T}\langle H_{0} \Phi\rangle_{\bf Q}^{0}\Bigr )
\nonumber\\
&=&\langle H_{0}\rangle_{\bf Q}^{0}
+\langle \Phi\rangle_{\bf Q}^{0}
-\frac{1}{k_{B}T}\Bigl (
\langle H_{0} \Phi\rangle_{\bf Q}^{0}
-\langle H_{0}\rangle_{\bf Q}^{0}
\langle \Phi\rangle_{\bf Q}^{0}\Bigr ).
\end{eqnarray}
Our purpose now is to determine all terms on the
right-hand side of Eq. (17) separately.
To find the average of the Gaussian Hamiltonian
$\langle H_{0}\rangle_{\bf Q}^{0}$,
we insert expression (1) into Eq. (11) and introduce
an explicit dependence of the Green function $G_{0}$
on segment length $b_{0}$
\begin{equation}
G_{0}(b_{0},{\bf Q})=
\int_{{\bf r}(0)={\bf 0}}^{{\bf r}(L)={\bf Q}}
\exp \biggl [-\frac{3}{2b_{0}}
\int_{0}^{L}\Bigl (\frac{d{\bf r}}{ds}(s)\Bigr )^{2} ds
\biggr ] {\cal D} [{\bf r}(s)].
\end{equation}
Differentiation of Eq. (18) with respect to $b_{0}$ implies that
\begin{eqnarray*}
\frac{\partial G_{0}}{\partial b_{0}}(b_{0},{\bf Q})
&=& \int_{{\bf r}(0)={\bf 0}}^{{\bf r}(L)={\bf Q}}
\biggl (\frac{3}{2b_{0}^{2}}
\int_{0}^{L}\Bigl (\frac{d{\bf r}}{ds}(s)\Bigr )^{2} ds\biggr )
\exp \biggl [-\frac{3}{2b_{0}}
\int_{0}^{L}\Bigl (\frac{d{\bf r}}{ds}(s)\Bigr )^{2} ds
\biggr ] {\cal D} [{\bf r}(s)]
\nonumber\\
&=& \frac{1}{k_{B}T b_{0}}
\int_{{\bf r}(0)={\bf 0}}^{{\bf r}(L)={\bf Q}}
H_{0}({\bf r}(s))
\exp \biggl [-\frac{H_{0}({\bf r}(s))}{k_{B}T}
\biggr ] {\cal D} [{\bf r}(s)]
\nonumber\\
&=& \frac{1}{k_{B}T b_{0}}\langle H_{0}\rangle_{\bf Q}^{0}
G_{0}(b_{0},{\bf Q}).
\end{eqnarray*}
It follows from this relation that
\begin{equation}
\langle H_{0}\rangle_{\bf Q}^{0}
=\frac{k_{B}Tb_{0}}{G_{0}(b_{0},{\bf Q})}
\frac{\partial G_{0}}{\partial b_{0}}(b_{0},{\bf Q}).
\end{equation}
Substitution of expression (8) with $b^{2}=b_{0}L$
into Eq. (19) yields
\begin{equation}
\langle H_{0}\rangle_{\bf Q}^{0}
=\frac{3k_{B}T}{2}\Bigl (\frac{Q^{2}}{b^{2}}-1\Bigr ).
\end{equation}
According to the definition of $\langle\Phi\rangle_{\bf Q}^{0}$
and Eq. (1), we have
\[
\langle\Phi\rangle_{\bf Q}^{0}G_{0}(b_{0},{\bf Q})=
\int_{{\bf r}(0)={\bf 0}}^{{\bf r}(L)={\bf Q}}
\Phi({\bf r}(s))\exp \biggl [-\frac{3}{2b_{0}}
\int_{0}^{L}\Bigl (\frac{d{\bf r}}{ds}(s)\Bigr )^{2} ds
\biggr ] {\cal D} [{\bf r}(s)],
\]
where an explicit dependence of $\langle\Phi\rangle_{\bf Q}^{0}$
on $b_{0}$ is omitted for brevity.
Differentiation of this equality with respect to $b_{0}$
results in
\begin{eqnarray}
\frac{\partial }{\partial b_{0}}
\Bigl (\langle\Phi\rangle_{\bf Q}^{0}G_{0}(b_{0},{\bf Q})\Bigr )
&=& \int_{{\bf r}(0)={\bf 0}}^{{\bf r}(L)={\bf Q}}
\biggl (\frac{3}{2b_{0}^{2}}
\int_{0}^{L}\Bigl (\frac{d{\bf r}}{ds}(s)\Bigr )^{2} ds\biggr )
\Phi({\bf r}(s))
\nonumber\\
&&\times
\exp \biggl [-\frac{3}{2b_{0}}
\int_{0}^{L}\Bigl (\frac{d{\bf r}}{ds}(s)\Bigr )^{2} ds
\biggr ] {\cal D} [{\bf r}(s)]
\nonumber\\
&=& \frac{1}{k_{B}T b_{0}}
\int_{{\bf r}(0)={\bf 0}}^{{\bf r}(L)={\bf Q}}
H_{0}({\bf r}(s))\Phi({\bf r}(s))
\exp \biggl [-\frac{H_{0}({\bf r}(s))}{k_{B}T}
\biggr ] {\cal D} [{\bf r}(s)]
\nonumber\\
&=& \frac{1}{k_{B}T b_{0}}\langle H_{0}\Phi\rangle_{\bf Q}^{0}
G_{0}(b_{0},{\bf Q}).
\end{eqnarray}
It follows from Eq. (21) that
\begin{eqnarray}
\langle H_{0}\Phi\rangle_{\bf Q}^{0}
&=&
\frac{k_{B}T b_{0}}{G_{0}(b_{0},{\bf Q})}
\frac{\partial }{\partial b_{0}}
\Bigl (\langle\Phi\rangle_{\bf Q}^{0}
G_{0}(b_{0},{\bf Q})\Bigr )
\nonumber\\
&=& k_{B}Tb_{0} \frac{\partial
\langle\Phi\rangle_{\bf Q}^{0}}{\partial b_{0}}
+\langle\Phi\rangle_{\bf Q}^{0}
\frac{k_{B}T b_{0}}{G_{0}(b_{0},{\bf Q})}
\frac{\partial G_{0}}{\partial b_{0}}
(b_{0},{\bf Q}).
\end{eqnarray}
Combination of Eqs. (19) and (22) implies that
\begin{equation}
\langle H_{0}\Phi\rangle_{\bf Q}^{0}
-\langle H_{0}\rangle_{\bf Q}^{0}
\langle \Phi \rangle_{\bf Q}^{0}
=k_{B}Tb_{0} \frac{\partial
\langle\Phi\rangle_{\bf Q}^{0}}{\partial b_{0}}.
\end{equation}
Inserting expressions (20) and (23) into Eq. (17),
we arrive at the formula
\begin{equation}
\tilde{\Psi}({\bf Q}) =
\frac{3k_{B}T}{2} \Bigl (\frac{Q^{2}}{b^{2}}-1\Bigr )
+\langle \Phi\rangle_{\bf Q}^{0}
-b_{0} \frac{\partial
\langle\Phi\rangle_{\bf Q}^{0}}{\partial b_{0}}.
\end{equation}
According to Eq. (24), to find the free energy of
a flexible chain with weak segment interactions,
we need to determine the energy of interactions
averaged with respect to the Gaussian Hamiltonian
$\langle\Phi\rangle_{\bf Q}^{0}$ only.
To calculate this quantity, a perturbative technique is applied.

\section{Correlation functions}

Our aim now is to derive an explicit expression
for the average of an exponential function of ${\bf r}(s)$
when the averaging is performed with respect to
the Hamiltonian $H_{0}$.
We begin with the perturbed Green function
\begin{eqnarray}
G^{\bf p}_{0}({\bf Q}) &=&
\int_{{\bf r}(0)={\bf 0}}^{{\bf r}(L)={\bf Q}}
\exp \biggl [-\frac{H_{0}({\bf r}(s))}{k_{\rm B}T}
+\int_{0}^{L} {\bf p}(s)\cdot \Bigl (
{\bf r}(s) -{\bf r}^{\circ}(s)\Bigr ) ds \biggr ]
{\cal D} [{\bf r}(s)]
\nonumber\\
&=& \langle \exp \Bigl (
\int_{0}^{L} {\bf p}(s)\cdot {\bf r}^{\prime} (s) ds\Bigr )
\rangle_{\bf Q}^{0} G_{0}({\bf Q}),
\end{eqnarray}
where ${\bf p}(s)$ is a smooth vector function determined on
$[0,L]$,
\begin{equation}
{\bf r}^{\circ}(s)={\bf Q}\frac{s}{L}
\end{equation}
is the ``classical" path for a Gaussian chain,
and
\begin{equation}
{\bf r}^{\prime}(s)={\bf r}(s)-{\bf r}^{\circ}(s).
\end{equation}
Substitution of Eqs. (26) and (27) into Eqs. (11) and (25)
results in
\begin{equation}
G^{\bf p}_{0}({\bf Q}) = \exp \Bigl (-\frac{3Q^{2}}{2b^{2}}\Bigr )
\bar{G}^{\bf p}{\bf Q}),
\qquad
G_{0}({\bf Q}) = \exp \Bigl (-\frac{3Q^{2}}{2b^{2}}\Bigr )
\bar{G}({\bf Q}),
\end{equation}
where
\begin{eqnarray}
\bar{G}^{\bf p}({\bf Q}) &=& \int
\exp \biggl [-\frac{3}{2b_{0}}
\int_{0}^{L} \Bigl (\frac{d{\bf r}^{\prime}}{ds}(s)\Bigr )^{2} ds
+\int_{0}^{L} {\bf p}(s)\cdot {\bf r}^{\prime}(s) ds \biggr ]
{\cal D} [{\bf r}^{\prime}(s)] ,
\nonumber\\
\bar{G}({\bf Q}) &=& \int
\exp \biggl [-\frac{3}{2b_{0}}
\int_{0}^{L} \Bigl (\frac{d{\bf r}^{\prime}}{ds}(s)\Bigr )^{2} ds
\biggr ] {\cal D} [{\bf r}^{\prime}(s)] ,
\end{eqnarray}
and the path integrals are calculated over
all curves ${\bf r}^{\prime}(s)$ that satisfy the
boundary conditions
\begin{equation}
{\bf r}^{\prime}(0)={\bf r}^{\prime}(L)={\bf 0}.
\end{equation}
Setting
\begin{equation}
{\bf r}^{\prime}(s)={\bf r}_{0}(s)+{\bf R}(s),
\end{equation}
where the functions ${\bf r}_{0}(s)$
and ${\bf R}(s)$ obey the zero boundary conditions
\begin{equation}
{\bf r}_{0}(0)={\bf r}_{0}(L)={\bf 0},
\qquad
{\bf R}(0)={\bf R}(L)={\bf 0}.
\end{equation}
we transform the expression in the square brackets in
the first equality in Eq. (29) as follows:
\begin{eqnarray}
A & \equiv& -\frac{3}{2b_{0}}
\int_{0}^{L} \Bigl (\frac{d{\bf r}^{\prime}}{ds}(s)\Bigr )^{2} ds
+\int_{0}^{L} {\bf p}(s)\cdot {\bf r}^{\prime}(s) ds
\nonumber\\
&=& -\frac{3}{2b_{0}} \int_{0}^{L} \Bigl [
\Bigl (\frac{d{\bf r}_{0}}{ds}(s)\Bigr )^{2}
+\Bigl (\frac{d{\bf R}}{ds}(s)\Bigr )^{2}
+2\frac{d{\bf r}_{0}}{ds}(s)\cdot \frac{d{\bf R}}{ds}(s)
\Bigr ] ds
\nonumber\\
&& +\int_{0}^{L} \Bigl [ {\bf p}(s)\cdot {\bf r}_{0}(s)
+{\bf p}(s)\cdot {\bf R}(s)\Bigr ] ds.
\end{eqnarray}
Integration by parts with the use of Eq. (32) yields
\[
\int_{0}^{L} \frac{d{\bf r}_{0}}{ds}(s)\cdot
\frac{d{\bf R}}{ds}(s) ds
=-\int_{0}^{L} {\bf r}_{0}(s)
\cdot \frac{d^{2}{\bf R}}{ds^{2}}(s) ds.
\]
It follows from this equality and Eq. (33) that
\begin{eqnarray*}
A &=& -\frac{3}{2b_{0}} \int_{0}^{L} \Bigl [
\Bigl (\frac{d{\bf r}_{0}}{ds}(s)\Bigr )^{2}
+\Bigl (\frac{d{\bf R}}{ds}(s)\Bigr )^{2} \Bigr ] ds
+\int_{0}^{L} \Bigl [ \frac{3}{b_{0}}
\frac{d^{2}{\bf R}}{ds^{2}}(s)+{\bf p}(s)\Bigr ]
\cdot {\bf r}_{0}(s) ds
\nonumber\\
&&
+\int_{0}^{L} {\bf p}(s)\cdot {\bf R}(s)ds.
\end{eqnarray*}
Assuming the function ${\bf R}(s)$ to satisfy the
differential equation
\begin{equation}
\frac{3}{b_{0}}\frac{d^{2}{\bf R}}{ds^{2}}(s)+{\bf p}(s)
=0,
\end{equation}
we find that
\begin{eqnarray}
A = -\frac{3}{2b_{0}} \int_{0}^{L} \Bigl [
\Bigl (\frac{d{\bf r}_{0}}{ds}(s)\Bigr )^{2}
+\Bigl (\frac{d{\bf R}}{ds}(s)\Bigr )^{2} \Bigr ] ds
+\int_{0}^{L} {\bf p}(s)\cdot {\bf R}(s)ds.
\end{eqnarray}
The solution of Eq. (34) with boundary conditions (32)
reads
\begin{equation}
{\bf R}(s) =\frac{b_{0}}{3} \int_{0}^{L}
D(s,s^{\prime}){\bf p}(s^{\prime}) ds^{\prime},
\end{equation}
where
\begin{equation}
D(s,s^{\prime})=s^{\prime}\Bigl (1-\frac{s}{L}\Bigr )
\quad (s\geq s^{\prime}),
\qquad
D(s,s^{\prime})=s\Bigl (1-\frac{s^{\prime}}{L}\Bigr )
\quad (s\leq s^{\prime}) .
\end{equation}
Integrating by parts the second term in the square brackets
in Eq. (35) and using Eqs. (32), (34) and (36), we obtain
\begin{eqnarray*}
\int_{0}^{L} \Bigl (\frac{d{\bf R}}{ds}(s)\Bigr )^{2} ds
&=&-\int_{0}^{L} {\bf R}(s)\cdot \frac{d^{2}{\bf R}}{ds^{2}}(s)
ds
=\frac{b_{0}}{3} \int_{0}^{L} {\bf p}(s) \cdot {\bf R}(s)
ds
\nonumber\\
&=& \Bigl (\frac{b_{0}}{3}\Bigr )^{2}
\int_{0}^{L}\int_{0}^{L}D(s,s^{\prime})
{\bf p}(s)\cdot {\bf p}(s^{\prime})ds ds^{\prime}.
\end{eqnarray*}
Substitution of this expression and Eq. (36) into Eq. (35)
results in
\begin{equation}
A = -\frac{3}{2b_{0}} \int_{0}^{L}
\Bigl (\frac{d{\bf r}_{0}}{ds}(s)\Bigr )^{2} ds
+\frac{b_{0}}{6} \int_{0}^{L}\int_{0}^{L}D(s,s^{\prime})
{\bf p}(s)\cdot {\bf p}(s^{\prime})ds ds^{\prime}.
\end{equation}
Bearing in mind that the last term in expression (38) can be
taken away from the path integral, we conclude from Eqs. (28),
(29) and (38) that
\[
G^{\bf p}_{0}({\bf Q})=G_{0}({\bf Q})\exp \biggl [
\frac{b_{0}}{6} \int_{0}^{L}\int_{0}^{L}
D(s,s^{\prime}) {\bf p}(s)\cdot {\bf p}(s^{\prime})
ds ds^{\prime} \biggr ].
\]
Comparison of this equality with Eq. (25) implies that
\begin{equation}
\langle \exp \Bigl [ \int_{0}^{L}
{\bf p}(s)\cdot {\bf r}^{\prime}(s) ds \Bigr ]
\rangle_{\bf Q}^{0}
=\exp \biggl [ \frac{b_{0}}{6} \int_{0}^{L}\int_{0}^{L}
D(s,s^{\prime}) {\bf p}(s)\cdot {\bf p}(s^{\prime})
ds ds^{\prime} \biggr ].
\end{equation}
Equation (39) serves as the basic tool for the analysis
of correlations between the radius vectors ${\bf r}(s)$
at various points $s\in [0,L]$ of a chain.
Setting
\[
{\bf p}(s)={\bf q}\delta (s-t_{1})-{\bf q}
\delta (s-t_{2}),
\]
where ${\bf q}$ is an arbitrary vector,
and $t_{1}, t_{2}\in (0,L)$ are arbitrary points,
we find from Eq. (39) that
\begin{equation}
\langle \exp \Bigl [ {\bf q}\cdot
\Bigl ( {\bf r}^{\prime}(t_{1})-
{\bf r}^{\prime}(t_{2})\Bigr )\Bigr ]
\rangle_{\bf Q}^{0}
=\exp \Bigl (\frac{b_{0}q^{2}}{6} \Delta (t_{1},t_{2})
\Bigr ),
\end{equation}
where
\begin{equation}
\Delta(t_{1},t_{2})= D(t_{1},t_{1})
-2D(t_{1},t_{2})+D(t_{2},t_{2}).
\end{equation}
In particular, when ${\bf q}=-\imath {\bf k}$, where
$\imath=\sqrt{-1}$, and ${\bf k}$ is a real vector,
Eq. (40) reads
\[
\langle \exp \Bigl [ -\imath {\bf k}\cdot
\Bigl ( {\bf r}^{\prime}(t_{1})
-{\bf r}^{\prime}(t_{2})\Bigr )\Bigr ] \rangle_{\bf Q}^{0}
=\exp \Bigl (-\frac{b_{0}k^{2}}{6} \Delta (t_{1},t_{2})\Bigr ).
\]
Replacing ${\bf r}^{\prime}$ by ${\bf r}$ in accord
with Eqs. (26) and (27), we obtain
\begin{eqnarray}
\langle \exp \Bigl [-\imath {\bf k}\cdot
\Bigl ( {\bf r}(t_{2})-{\bf r}(t_{1})\Bigr )\Bigr ]
\rangle_{\bf Q}^{0}
&=& \langle \exp \Bigl [ -\imath {\bf k}\cdot
\Bigl ( {\bf r}^{\prime}(t_{1})
-{\bf r}^{\prime}(t_{2})\Bigr )\Bigr ] \rangle_{\bf Q}^{0}
\exp \Bigl (-\imath {\bf k}\cdot {\bf Q}
\frac{t_{2}-t_{1}}{L}\Bigr )
\nonumber\\
&=& \exp \Bigl (-\frac{b_{0}k^{2}}{6} \Delta (t_{1},t_{2})
-\imath {\bf k}\cdot {\bf Q} \frac{t_{2}-t_{1}}{L}\Bigr ).
\end{eqnarray}
Our aim now is to apply Eq. (42) in order to calculate
the average energy of segment interactions
$\langle \Phi\rangle_{\bf Q}^{0}$ for the functional
$\Phi({\bf r})$ given by Eq. (3).

\section{The average energy of segment interactions}

We introduce the Fourier transform of the delta-function
by the formula
\begin{equation}
\delta({\bf r})=\frac{1}{(2\pi)^{3}}
\int \exp ( -\imath {\bf k}\cdot {\bf r})d{\bf k},
\end{equation}
combine Eqs. (3) and (43), and find that
in a spherical coordinate frame $\{ k,\phi,\theta \}$,
whose ${\bf e}_{3}$ vector is directed along
the end-to-end vector ${\bf Q}$, the functional $\Phi$ reads
\begin{eqnarray*}
\Phi({\bf r}) &=& \frac{v_{0}}{2L^{2}(2\pi)^{3}}
\int_{0}^{\infty} k^{2} d k
\int_{0}^{2\pi} d\phi\int_{0}^{\pi} \sin \theta d\theta
\int_{0}^{L} ds\int_{0}^{L}
\exp \Bigl [-\imath {\bf k}\cdot \Bigl (
{\bf r}(s)-{\bf r}(s^{\prime})\Bigr )\Bigr ]
ds^{\prime}
\nonumber\\
&=& \frac{v_{0}}{L^{2}(2\pi)^{3}}
\int_{0}^{\infty} k^{2} d k
\int_{0}^{2\pi} d\phi\int_{0}^{\pi} \sin \theta d\theta
\int_{0}^{L} ds\int_{0}^{s}
\exp \Bigl [-\imath {\bf k}\cdot \Bigl (
{\bf r}(s)-{\bf r}(s^{\prime})\Bigr )\Bigr ]
ds^{\prime}.
\end{eqnarray*}
It follows from this relation and Eq. (42) that
\begin{eqnarray*}
\langle \Phi\rangle_{\bf Q}^{0}
&=& \frac{v_{0}}{L^{2}(2\pi)^{3}}
\int_{0}^{\infty} k^{2} d k
\int_{0}^{2\pi} d\phi\int_{0}^{\pi} \sin \theta d\theta
\int_{0}^{L} ds
\nonumber\\
&&\times
\int_{0}^{s} \exp \Bigl (-\frac{b_{0}k^{2}}{6}
\Delta (s,s^{\prime}) -\imath k Q \cos \theta
\frac{s-s^{\prime}}{L} \Bigr ) ds^{\prime}.
\end{eqnarray*}
Equations (37) and (41) imply that for any $s\geq s^{\prime}$,
\[
\Delta (s,s^{\prime})=(s-s^{\prime})-\frac{1}{L}
(s-s^{\prime})^{2}.
\]
Using this equality, performing integration over $\phi$,
and setting $\tau=s-s^{\prime}$, we obtain
\begin{eqnarray*}
\langle \Phi\rangle_{\bf Q}^{0}
&=& \frac{v_{0}}{(2\pi L)^{2}}
\int_{0}^{\infty} k^{2} d k
\int_{0}^{\pi} \sin \theta d\theta
\int_{0}^{L} ds
\int_{0}^{s} \exp \Bigl [-\frac{b_{0}k^{2}}{6}
\Bigl (\tau-\frac{\tau^{2}}{L}\Bigr )
-\imath k Q \cos \theta \frac{\tau}{L} \Bigr ] d\tau
\nonumber\\
&=& \frac{v_{0}}{(2\pi L)^{2}}
\int_{0}^{\infty} k^{2} d k
\int_{0}^{\pi} \sin \theta d\theta
\int_{0}^{L} (L-\tau) \exp \Bigl [-\frac{b_{0}k^{2}}{6}
\Bigl (\tau-\frac{\tau^{2}}{L}\Bigr )
-\imath k Q \cos \theta \frac{\tau}{L} \Bigr ] d\tau,
\end{eqnarray*}
where we changed the order of integration over $s$ and
$\tau$ and integrated over $s$ explicitly.
Introducing the notation $x=\cos \theta$ and $t=\tau/L$,
we find that
\[
\langle \Phi\rangle_{\bf Q}^{0}
= \frac{v_{0}}{(2\pi )^{2}}
\int_{0}^{\infty} k^{2} d k
\int_{-1}^{1} dx
\int_{0}^{1} \exp \Bigl [-\frac{(bk)^{2}}{6}t
(1-t) -\imath k Q x t \Bigr ] (1-t) d t.
\]
Bearing in mind that
\[
\int_{-1}^{1} \exp (-\imath k Q x t ) d x
=\frac{2\sin(kQt)}{kQt},
\]
we arrive at the formula
\begin{equation}
\langle \Phi\rangle_{\bf Q}^{0}
= \frac{v_{0}}{2\pi^{2}}
\int_{0}^{\infty} M_{0}(kQ,k) k^{2} d k ,
\end{equation}
where
\[
M_{0}(a,k)=\int_{0}^{1} m(at)\exp
\Bigl [-\frac{(bk)^{2}}{6}t(1-t) \Bigr ] (1-t) d t
\]
and
\[
m(x)= \frac{\sin x}{x} .
\]
Differentiation of Eq. (44) with respect to $b_{0}$
implies that
\begin{equation}
b_{0}\frac{\partial
\langle \Phi\rangle_{\bf Q}^{0}}{\partial b_{0}}
= -\frac{v_{0}}{2\pi^{2}}
\int_{0}^{\infty} M_{1}(kQ,k) k^{2} d k ,
\end{equation}
where
\[
M_{1}(a,k)=\frac{(bk)^{2}}{6}
\int_{0}^{1} m(at)\exp \Bigl [-\frac{(bk)^{2}}{6}
t(1-t) \Bigr ] t(1-t)^{2} d t.
\]
Substitution of expressions (44) and (45) into Eq. (24)
yields
\begin{equation}
\tilde{\Psi}({\bf Q})=
\frac{3k_{B}T}{2} \Bigl (\frac{Q^{2}}{b^{2}}-1\Bigr )
+\frac{v_{0}}{2\pi^{2}}
\int_{0}^{\infty} M(kQ,k) k^{2} d k ,
\end{equation}
where
\[
M(a,k)=\int_{0}^{1} m(at)
\exp \Bigl [-\frac{(bk)^{2}}{6}t(1-t) \Bigr ]
\biggl (1+\frac{(bk)^{2}}{6}t(1-t)\biggr )
(1-t)d t.
\]
Changing the order of integration in Eq. (46),
we obtain
\begin{equation}
\tilde{\Psi}({\bf Q})=
\frac{3k_{B}T}{2} \Bigl (\frac{Q^{2}}{b^{2}}-1\Bigr )
+\frac{v_{0}}{2\pi^{2}Q} \int_{0}^{1}
\frac{1-t}{t}J(t,Q)dt,
\end{equation}
where
\[
J(t,Q) = \int_{0}^{\infty}
\exp \Bigl [-\frac{(bk)^{2}}{6}t(1-t) \Bigr ]
\biggl (1+\frac{(bk)^{2}}{6}t(1-t)\biggr )
\sin(kQt)k dk.
\]
Bearing in mind that the function under the integral is even,
we present this equality in the form
\begin{equation}
J(t,Q) = \frac{1}{2} \int_{-\infty}^{\infty}
\exp \Bigl [-\frac{(bk)^{2}}{6}t(1-t) \Bigr ]
\biggl (1+\frac{(bk)^{2}}{6}t(1-t)\biggr )
\sin(kQt)k dk.
\end{equation}
The integral in Eq. (48) is calculated
with the help of the formulas
\begin{eqnarray*}
&& \int_{-\infty}^{\infty} \exp \Bigl (-\frac{\alpha k^{2}}{2}\Bigr )
\sin (\beta k) k dk
=\sqrt{\frac{2\pi}{\alpha}} \frac{\beta}{\alpha}
\exp \Bigl (-\frac{\beta^{2}}{2\alpha} \Bigr ),
\nonumber\\
&& \int_{-\infty}^{\infty} \exp \Bigl (-\frac{\alpha k^{2}}{2}\Bigr )
\frac{\alpha k^{2}}{2} \sin (\beta k) k dk
=\frac{1}{2} \sqrt{\frac{2\pi}{\alpha}} \frac{\beta}{\alpha}
\exp \Bigl (-\frac{\beta^{2}}{2\alpha} \Bigr )
\Bigl (3-\frac{\beta^{2}}{\alpha}\Bigr ),
\end{eqnarray*}
that are fulfilled for any $\alpha>0$.
Combination of these relations with Eq. (48) implies that
\[
J(t,Q)=\frac{3Q\sqrt{6\pi}}{4b^{3}\sqrt{t(1-t)^{3}}}
\Bigl (5-\frac{3Q^{2}t}{b^{2}(1-t)}\Bigr )
\exp \Bigl (-\frac{3Q^{2}t}{2b^{2}(1-t)}\Bigr ).
\]
Inserting this expression into Eq. (47), we find that
\begin{equation}
\tilde{\Psi}({\bf Q})= \frac{3k_{B}T}{2} \biggl [
\Bigl (\frac{Q^{2}}{b^{2}}-1\Bigr )
+\frac{v\sqrt{6\pi}}{4\pi^{2}b^{3}} \Bigl (5 R_{1}(Q)
-\frac{3Q^{2}}{b^{2}} R_{2}(Q)\Bigr ) \biggr ],
\end{equation}
where $v=v_{0}/(k_{B}T)$, and
\begin{eqnarray}
R_{1}(Q) &=& \int_{0}^{1}
\exp \Bigl (-\frac{3Q^{2}t}{2b^{2}(1-t)}\Bigr )
\frac{dt}{\sqrt{t^{3}(1-t)}},
\nonumber\\
R_{2}(Q) &=& \int_{0}^{1}
\exp \Bigl (-\frac{3Q^{2}t}{2b^{2}(1-t)}\Bigr )
\frac{dt}{\sqrt{t(1-t)^{3}}}.
\end{eqnarray}
The function $R_{2}(Q)$ reads
\begin{eqnarray}
R_{2}(Q)&=& \int_{0}^{\infty} \exp
\Bigl (-\frac{3Q^{2}\tau}{2b^{2}}\Bigr )
\frac{d\tau}{\sqrt{\tau}}
=2\int_{0}^{\infty} \exp \Bigl (-\frac{3Q^{2}s^{2}}{2b^{2}}
\Bigr )ds
\nonumber\\
&=& \frac{2b}{Q\sqrt{3}}\int_{0}^{\infty} \exp
\Bigl (-\frac{z^{2}}{2} \Bigr ) dz
=\frac{b}{Q}\sqrt{\frac{2\pi}{3}},
\end{eqnarray}
where we used the following variables:
$\tau=t/(1-t)$, $s=\sqrt{\tau}$, and
$z=Qs\sqrt{3}/b$.
The first integral in Eq. (50) is presented in the form
\begin{equation}
R_{1}(Q) = R_{0} -\int_{0}^{1}\biggl [1-
\exp \Bigl (-\frac{3Q^{2}t}{2b^{2}(1-t)}\Bigr )\biggr ]
\frac{dt}{\sqrt{t^{3}(1-t)}},
\end{equation}
where $R_{0}$ is independent of $Q$.
According to the remark at the end of Section 2,
the additive constant $R_{0}$ does not affect
the increment of free energy $\Delta\tilde{\Psi}$
and the distribution function $\tilde{p}$,
and we do not calculate this quantity.
Some concern may arise regarding $R_{0}$,
because the integral in the formula for $R_{0}$
diverges.
This divergence does not affect, however, the free energy.
To avoid it, one can replace Eq. (43) for the delta-function
by its regularization,
\[
\delta({\bf r})=\frac{1}{(2\pi)^{3}}
\int U(k) \exp ( -\imath {\bf k}\cdot {\bf r})d{\bf k},
\]
where
\[
U(k)=1
\quad
(0\leq k\leq k_{\ast}),
\qquad
U(k)=0,
\quad
(k>k_{\ast})
\]
with some $k_{\ast}\gg 1$, find the free energy of a
flexible chain with the regularized potential of
segment interactions,
disregard the additive constant in the
expression for $\tilde{\Psi}({\bf Q})$,
and, afterwards, take the limit at $k_{\ast}\to \infty$.
We do not dwell on detailed transformations,
because they cause unnecessary complications of
the analysis without influence on the final result.

The term dependent on $Q$ in Eq. (52) is transformed
as follows:
\begin{eqnarray*}
&& \int_{0}^{1}\biggl [1-
\exp \Bigl (-\frac{3Q^{2}t}{2b^{2}(1-t)}\Bigr )\biggr ]
\frac{dt}{\sqrt{t^{3}(1-t)}}
=\int_{0}^{\infty}\Bigl [ 1-\exp
\Bigl (-\frac{3Q^{2}\tau}{2b^{2}}\Bigr )\Bigr ]
\tau^{-\frac{3}{2}} d\tau
\nonumber\\
&& =\frac{Q}{b}\sqrt{\frac{3}{2}}\int_{0}^{\infty}
\Bigl (1-\exp (-s)\Bigr )s^{-\frac{3}{2}}ds,
\end{eqnarray*}
where we set $\tau=t/(1-t)$ and $s=3Q^{2}\tau/(2b^{2})$.
Integration by parts results in
\[
\int_{0}^{\infty}
\Bigl (1-\exp (-s)\Bigr )s^{-\frac{3}{2}}ds
=2\int_{0}^{\infty} y^{-\frac{1}{2}}\exp (-y) dy
=2\Gamma(\frac{1}{2})=2\sqrt{\pi},
\]
where $\Gamma(x)$ is the Euler gamma-function.
Combining these relations, we find that
\begin{equation}
R_{1}(Q) = R_{0}- \frac{Q}{b}\sqrt{6\pi}.
\end{equation}
Substituting expressions (51) and (53) into Eq. (49)
and neglecting additive constants, we arrive at
the formula
\begin{equation}
\tilde{\Psi}({\bf Q})=\frac{3k_{B}T}{2}\Bigl (
\frac{Q^{2}}{b^{2}}-\varepsilon \frac{Q}{b}\Bigr ),
\end{equation}
where
\[
\varepsilon=\frac{9v}{\pi b^{3}}
\]
is the dimensionless strength of excluded volume
interactions.
It follows from Eqs. (14) and (54) that
\begin{equation}
\tilde{p}({\bf Q})=p_{0}\exp \Bigl [-\frac{3}{2}
\Bigl (\frac{Q^{2}}{b^{2}}-\varepsilon \frac{Q}{b}\Bigr )
\Bigr ],
\end{equation}
where the pre-factor $p_{0}$ is determined from the
normalization condition
\begin{equation}
p_{0}=\frac{1}{4\pi}\biggl \{ \int_{0}^{\infty}
\exp \Bigl [-\frac{3}{2}
\Bigl (\frac{Q^{2}}{b^{2}}-\varepsilon \frac{Q}{b}\Bigr )
\Bigr ]Q^{2} dQ\biggr \}^{-1}.
\end{equation}
For weak excluded-volume interactions
($\varepsilon\ll 1$),
the integral in Eq. (56) is given by
\begin{eqnarray*}
\int_{0}^{\infty} \exp \Bigl [-\frac{3}{2}
\Bigl (\frac{Q^{2}}{b^{2}}-\varepsilon \frac{Q}{b}\Bigr )
\Bigr ]Q^{2} dQ
&=& \frac{b^{3}}{3\sqrt{3}} \biggl [ \int_{0}^{\infty} \exp
\Bigl (-\frac{z^{2}}{2} \Bigr ) z^{2}dz
+\frac{\varepsilon\sqrt{3}}{2}\int_{0}^{\infty}
\Bigl (-\frac{z^{2}}{2} \Bigr ) z^{3}dz \biggr ]
\nonumber\\
&=& \frac{b^{3}}{3\sqrt{3}}\Bigl (\sqrt{\frac{\pi}{2}}
+\varepsilon\sqrt{3}\Bigr ),
\end{eqnarray*}
where $z=Q\sqrt{3}/b$.
This relation together with Eq. (56) implies that
\begin{equation}
p_{0}=\Bigl (\frac{3}{2\pi b^{2}}\Bigr )^{\frac{3}{2}}
\Bigl (1-\varepsilon \sqrt{\frac{6}{\pi}}\Bigr )
\end{equation}
in the first approximation with respect to $\varepsilon$.
Inserting Eqs. (54) and (55) into Eq. (15)
and introducing a spherical coordinate
frame $\{ Q,\phi,\theta \}$, we obtain
\begin{eqnarray}
\tilde{W} &=& \frac{3k_{B}Tp_{0}}{2} \int_{0}^{\infty}
\exp \Bigl [ -\frac{3}{2}\Bigl (\frac{Q^{2}}{b^{2}}
-\varepsilon \frac{Q}{b}\Bigr )\Bigr ] Q^{2} dQ
\int_{0}^{2\pi} d\phi
\nonumber\\
&&\times
\int_{0}^{\pi} \Bigl [
\frac{1}{b^{2}} \Bigl ({\bf Q}\cdot {\bf C}\cdot {\bf Q}
-Q^{2}\Bigr )-\frac{\varepsilon}{b}\Bigl (
\sqrt{{\bf Q}\cdot {\bf C}\cdot {\bf Q}}-Q \Bigr )\Bigr ]
\sin \theta d\theta.
\end{eqnarray}
Denote by ${\bf i}_{m}$ the eigenvectors of the
the right Cauchy--Green tensor ${\bf C}$ and by $\lambda_{m}$
appropriate eigenvalues.
If the ${\bf e}_{3}$ vector of the spherical coordinate
frame is directed along the eigenvector ${\bf i}_{3}$,
the expression ${\bf Q}\cdot {\bf C}\cdot {\bf Q}$
reads
\[
{\bf Q}\cdot {\bf C}\cdot {\bf Q}=Q^{2}S,
\qquad
S=(\lambda_{1}\cos^{2}\phi
+\lambda_{2}\sin^{2}\phi)\sin^{2}\theta
+\lambda_{3}\cos^{2} \theta.
\]
Substitution of this relation into Eq. (58) results in
\begin{equation}
\tilde{W}=W_{1}-W_{2},
\end{equation}
where
\begin{eqnarray*}
W_{1} &=& \frac{3k_{B}Tp_{0}b^{3}}{2}
\int_{0}^{\infty} \exp \Bigl [
-\frac{3}{2} (z^{2}-\varepsilon z )
\Bigr ] z^{4} dz \int_{0}^{2\pi} d\phi
\nonumber\\
&&\times
\int_{0}^{\pi} \Bigl [
\Bigl ((\lambda_{1}\cos^{2}\phi
+\lambda_{2}\sin^{2}\phi)\sin^{2}\theta
+\lambda_{3}\cos^{2} \theta \Bigr )-1\Bigr ]
\sin \theta d\theta,
\nonumber\\
W_{2} &=& \frac{3\varepsilon k_{B}Tp_{0}b^{3}}{2}
\int_{0}^{\infty} \exp \Bigl [
-\frac{3}{2}(z^{2}-\varepsilon z)\Bigr ] z^{3} dz
\int_{0}^{2\pi} d\phi
\nonumber\\
&&\times
\int_{0}^{\pi} \Bigl [
\Bigl ((\lambda_{1}\cos^{2}\phi
+\lambda_{2}\sin^{2}\phi)\sin^{2}\theta
+\lambda_{3}\cos^{2} \theta \Bigr )^{\frac{1}{2}} -1\Bigr ]
\sin \theta d\theta ,
\end{eqnarray*}
and $z=Q/b$.
We calculate the integrals over $z$,
substitute expression (57) for the coefficient $p_{0}$,
disregards terms beyond the first order of smallness
with respect to $\varepsilon$, and find that
\begin{eqnarray}
W_{1} &=& \frac{3k_{B}T}{8\pi}\Bigl (1+\frac{\varepsilon}{3}
\sqrt{\frac{6}{\pi}}\Bigr ) \int_{0}^{2\pi} d\phi
\int_{0}^{\pi} \Bigl [
\Bigl ((\lambda_{1}\cos^{2}\phi
+\lambda_{2}\sin^{2}\phi)\sin^{2}\theta
+\lambda_{3}\cos^{2} \theta \Bigr )-1\Bigr ]
\sin \theta d\theta,
\nonumber\\
W_{2} &=& \frac{\varepsilon k_{B}T}{4\pi}\sqrt{\frac{6}{\pi}}
\int_{0}^{2\pi} d\phi
\int_{0}^{\pi} \Bigl [
\Bigl ((\lambda_{1}\cos^{2}\phi
+\lambda_{2}\sin^{2}\phi)\sin^{2}\theta
+\lambda_{3}\cos^{2} \theta \Bigr )^{\frac{1}{2}} -1\Bigr ]
\sin \theta d\theta .
\end{eqnarray}
Calculating the integrals over $\phi$ and $\theta$
in the first equality in Eq. (60), we arrive at the formula
\begin{equation}
W_{1}=\frac{k_{B}T}{2}\Bigl (1+\frac{\varepsilon}{3}
\sqrt{\frac{6}{\pi}}\Bigr )
(\lambda_{1}+\lambda_{2}+\lambda_{3}-3).
\end{equation}
According to Eqs. (9) and (61), at $\varepsilon=0$
(no intra-chain interactions),
$W_{1}$ coincides with the strain energy of a Gaussian chain.
Setting $x=\cos\theta$ in the other equality in Eq. (60)
and using the evenness of the function under the integral,
we find that
\begin{eqnarray}
W_{2}
&=& \frac{\varepsilon k_{B}T}{2\pi}\sqrt{\frac{6}{\pi}}
\int_{0}^{2\pi} d\phi
\int_{0}^{1} \Bigl [ \Bigl ((\lambda_{1}\cos^{2}\phi
+\lambda_{2}\sin^{2}\phi)(1-x^{2})
+\lambda_{3}x^{2} \Bigr )^{\frac{1}{2}} -1\Bigr ]d x
\nonumber\\
&=& \varepsilon k_{B}T\sqrt{\frac{6}{\pi}}
\biggl [ \frac{1}{2\pi} \int_{0}^{2\pi} d\phi
\int_{0}^{1} \Bigl ((\lambda_{1}\cos^{2}\phi
+\lambda_{2}\sin^{2}\phi)(1-x^{2})
+\lambda_{3}x^{2} \Bigr )^{\frac{1}{2}}dx -1\biggr ]
\nonumber\\
&=& \varepsilon k_{B}T\sqrt{\frac{6}{\pi}}
\biggl [ \frac{2}{\pi} \int_{0}^{\frac{\pi}{2}} d\phi
\int_{0}^{1} \Bigl ((\lambda_{1}\cos^{2}\phi
+\lambda_{2}\sin^{2}\phi)(1-x^{2})
+\lambda_{3}x^{2} \Bigr )^{\frac{1}{2}}dx -1\biggr ].
\end{eqnarray}
Although it is possible to develop an analytical
expression for the strain energy $W_{2}$
for an arbitrary three-dimensional deformation,
an appropriate formula is rather cumbersome,
and we do not present it for the sake of brevity.
We confine ourselves to a particular case of
axisymmetric deformation with
\begin{equation}
\lambda_{1}=\lambda_{2}=\lambda
\end{equation}
for two reasons: (i) under condition (63)
the governing relations remain relatively simple,
and (ii) deformation processes (63) are typical
for experiments on uniaxial and equi-biaxial extension
of elastomers.
Performing integration over $\phi$ in Eq. (62), we obtain
\[
W_{2}=\varepsilon k_{B}T\sqrt{\frac{6}{\pi}}
\biggl [ \int_{0}^{1} \Bigl (\lambda
+(\lambda_{3}-\lambda)x^{2} \Bigr )^{\frac{1}{2}}dx -1\biggr ].
\]
Calculation of the integral over $x$ implies that
\begin{eqnarray}
W_{2}&=&\frac{\varepsilon k_{B}T}{2}\sqrt{\frac{6}{\pi}}
\biggl (\sqrt{\lambda_{3}}+\frac{\lambda}
{\sqrt{\lambda_{3}-\lambda}}\ln \frac{\sqrt{\lambda_{3}}
+\sqrt{\lambda_{3}-\lambda}}{\sqrt{\lambda}}-2\biggr )
\quad
(\lambda_{3}>\lambda),
\nonumber\\
W_{2} &=& \frac{\varepsilon k_{B}T}{2}\sqrt{\frac{6}{\pi}}
\biggl (\sqrt{\lambda_{3}}
+\frac{\lambda}{\sqrt{\lambda-\lambda_{3}}}
\arcsin \sqrt{\frac{\lambda-\lambda_{3}}{\lambda}}
-2\biggr )
\quad
(\lambda_{3}<\lambda),
\nonumber\\
W_{2} &=& \varepsilon k_{B}T \sqrt{\frac{6}{\pi}}
\Bigl (\sqrt{\lambda_{3}}-1\Bigr)
\quad
(\lambda_{3}=\lambda).
\end{eqnarray}
To obtain the strain energy $W_{2}$ as a function of
the principal stretches $\lambda_{m}$,
the parameter $\lambda$ in Eq. (64) should be replaced
by $\frac{1}{2} (\lambda_{1}+\lambda_{2})$.

\section{Constitutive equations for a network of
flexible chains}

To develop stress--strain relations for a permanent
network of flexible chains with excluded volume
interactions, we adopt the following hypotheses:
(i) the motion of chains is affine, which means that
the deformation gradient ${\bf F}$ coincides with
the deformation gradient for macro-deformation,
and (ii) inter-chain interactions are accounted for by
using the incompressibility condition, which implies
that the strain energy of a network equals the sum
of strain energies of individual chains \cite{DE86}.
Denote by $M$ the number of chains per unit volume.
It follows from Eqs. (59), (61) and (62) that the strain
energy density (per unit volume of the network) reads
\begin{eqnarray}
\bar{W}(\lambda_{m}) &=& \frac{k_{B}TM}{2}\biggl \{
\Bigl (1+\frac{\varepsilon}{3} \sqrt{\frac{6}{\pi}}\Bigr )
(\lambda_{1}+\lambda_{2}+\lambda_{3}-3)
\nonumber\\
&& -2\varepsilon\sqrt{\frac{6}{\pi}}
\biggl [ \frac{2}{\pi} \int_{0}^{\frac{\pi}{2}} d\phi
\int_{0}^{1} \Bigl ((\lambda_{1}\cos^{2}\phi
+\lambda_{2}\sin^{2}\phi)(1-x^{2})
+\lambda_{3}x^{2} \Bigr )^{\frac{1}{2}}dx -1\biggr ]
\biggr \}.
\end{eqnarray}
The principal Cauchy stresses $\Sigma_{m}$ are expressed
in terms of the strain energy $\bar{W}$ as \cite{Dro96}
\begin{equation}
\Sigma_{m}=-P+\lambda_{m}
\frac{\partial \bar{W}}{\partial \lambda_{m}},
\end{equation}
where $P$ stands for pressure.
Formulas (65) and (66) provide the stress--strain
relations for an incompressible network of flexible chains
with weak excluded-volume interactions.

Our aim now is to apply these equations in order to
evaluate the effect of segment interactions on
the elastic response of a polymer network at uniaxial
tension (compression) and equi-biaxial tension.

\subsection{Uniaxial tension}

Uniaxial tension of an incompressible medium is described
by the formulas
\[
x_{1}=k^{-\frac{1}{2}}X_{1},
\qquad
x_{2}=k^{-\frac{1}{2}}X_{2},
\qquad
x_{3}=k X_{3},
\]
where $\{ X_{m}\}$  and $\{ x_{m} \}$ are Cartesian
coordinates in the reference and actual states,
respectively, and $k$ denotes elongation ratio.
The right Cauchy--Green tensor ${\bf C}$ is given by
\[
{\bf C}=k^{-1} ({\bf e}_{1}{\bf e}_{1}
+{\bf e}_{2}{\bf e}_{2})
+k^{2} {\bf e}_{3}{\bf e}_{3},
\]
where ${\bf e}_{m}$ are base vectors of the Cartesian
frame in the initial state, and its eigenvalues read
\begin{equation}
\lambda_{1}=\lambda_{2}=k^{-1},
\qquad
\lambda_{3}=k^{2}.
\end{equation}
Inserting expressions (67) into Eq. (65) and using Eq. (64),
we find that
\begin{eqnarray}
\bar{W} &=& \frac{k_{B}TM}{2}\biggl [
\Bigl (1+\frac{\varepsilon}{3} \sqrt{\frac{6}{\pi}}\Bigr )
\Bigl (k^{2}+\frac{2}{k}-3\Bigr )
-\varepsilon\sqrt{\frac{6}{\pi}}
\Bigl ( k+\frac{\ln(\sqrt{k^{3}}
+\sqrt{k^{3}-1})}{\sqrt{k(k^{3}-1)}}
-2\Bigr )\biggr ]
\qquad (k>1),
\nonumber\\
\bar{W} &=& \frac{k_{B}TM}{2}\biggl [
\Bigl (1+\frac{\varepsilon}{3} \sqrt{\frac{6}{\pi}}\Bigr )
\Bigl (k^{2}+\frac{2}{k}-3\Bigr )
-\varepsilon\sqrt{\frac{6}{\pi}}
\Bigl ( k+\frac{\arcsin \sqrt{1-k^{3}}}{\sqrt{k(1-k^{3})}}
-2\Bigr )\biggr ]
\qquad (k<1).
\end{eqnarray}
According to Eq. (66), at uniaxial tension (compression)
the longitudinal Cauchy stress $\Sigma$ is determined as
\[
\Sigma=\lambda_{3}\frac{\partial \bar{W}}{\partial \lambda_{3}}
-\lambda_{1}\frac{\partial \bar{W}}{\partial \lambda_{1}}
=\lambda_{3}\frac{\partial \bar{W}}{\partial \lambda_{3}}
-\frac{1}{2} \lambda_{1}\frac{\partial \bar{W}}{\partial \lambda}.
\]
Substitution of expressions (67) into this equality yields
\[
\Sigma=k\frac{d\bar{W}}{dk},
\]
which implies that the engineering tensile stress
$\sigma=\Sigma/k$ is given by
\begin{equation}
\sigma=\frac{d\bar{W}}{dk}.
\end{equation}
Combination of Eqs. (68) and (69) implies that
\begin{eqnarray}
\sigma &=& k_{B}TM \biggl [
\Bigl (1+\frac{\varepsilon}{3} \sqrt{\frac{6}{\pi}}\Bigr )
\Bigl (k-\frac{1}{k^{2}}\Bigr )
\nonumber\\
&& -\frac{\varepsilon}{2}\sqrt{\frac{6}{\pi}}
\Bigl ( 1+\frac{3\sqrt{k^{3}(k^{3}-1)}-(4k^{3}-1)\ln (\sqrt{k^{3}}
+\sqrt{k^{3}-1})}{2\sqrt{k^{3}(k^{3}-1)^{3}}}\Bigr )\biggr ]
\qquad
(k>1),
\nonumber\\
\sigma &=& k_{B}TM \biggl [
\Bigl (1+\frac{\varepsilon}{3} \sqrt{\frac{6}{\pi}}\Bigr )
\Bigl (k-\frac{1}{k^{2}}\Bigr )
-\frac{\varepsilon}{2}\sqrt{\frac{6}{\pi}}
\nonumber\\
&&\times
\Bigl ( 1-\frac{3\sqrt{k^{3}(1-k^{3})}
+(1-4k^{3})\arcsin \sqrt{1-k^{3}}}
{2\sqrt{k^{3}(1-k^{3})^{3}}}\Bigr )\biggr ]
\qquad
(k<1).
\end{eqnarray}
It is easy to check that $\lim_{k\to 1} \sigma(k)=0$,
which means that the reference state is stress-free.
Differentiation of the first equality in Eq. (70) with
respect to $k$ yields
\begin{eqnarray}
\frac{d\sigma}{dk} &=& k_{B}TM \biggl \{
\Bigl (1+\frac{\varepsilon}{3} \sqrt{\frac{6}{\pi}}\Bigr )
\Bigl (1+\frac{2}{k^{3}}\Bigr )
-\frac{3\varepsilon}{8k^{\frac{5}{2}}}\sqrt{\frac{6}{\pi}}
\nonumber\\
&&\times
\frac{1}{(k^{3}-1)^{\frac{5}{2}}}
\Bigl [ (8k^{6}+1)\ln (\sqrt{k^{3}}+\sqrt{k^{3}-1})
-(10k^{3}-1)\sqrt{k^{3}(k^{3}-1)} \Bigr ] \biggl \}.
\end{eqnarray}
The Young's modulus of a network is defined as
\[
E=\frac{d\sigma}{dk}\biggl |_{k=1}.
\]
Inserting expression (71) into this equality
and applying the L'Hospital rule to calculate
the limit at $k\to 1$, we find that
\begin{equation}
E=E_{0}\Bigl (1+\frac{\varepsilon}{15}
\sqrt{\frac{6}{\pi}}\Bigr ),
\end{equation}
where
\[
E_{0}=3 k_{B}TM
\]
stands for the Young's modulus of a network of Gaussian
chains.
According to Eq. (72), excluded-volume interactions induce
an increase in the elastic modulus of a network of flexible
chains which is proportional to the dimensionless strength
of segment interactions $\varepsilon$.

To assess the effect of intra-chain interactions on the
stress--strain diagrams, we perform numerical simulation
of Eqs. (70) at uniaxial tension and compression.
The results of numerical analysis are presented in
Figures 1 (tension) and 2 (compression),
where the reduced tensile stress
\[
\sigma_{\ast}=\frac{\sigma_{0}}{k-k^{-2}},
\]
is depicted versus elongation ratio $k$
(the Mooney--Rivlin plots).
Here
$\sigma_{0}=\sigma/(k_{B}TM)$ stands for
the dimensionless engineering tensile stress.
In these figures, the functions $\sigma_{\ast}(k)$
for a network of Gaussian chains are presented by horizontal
lines,
whereas appropriate dependencies for a network of flexible
chains with segment interactions demonstrate a monotonic
increase in tensile and compressive stresses
(it is worth noting that a weak minimum of $\sigma_{\ast}$
at compression observed in Figure 2 does not reflect
non-monotonicity of the dependence $\sigma(k)$;
the latter function decreases with $k$ in the
entire domain $k\in (0,1)$).
The results plotted in Figures 1 and 2 indicate
that excluded-volume interactions cause the growth of
stiffness of a polymer network, in agreement with Eq. (72)
that determines the elastic modulus at small strains.

Our aim now is to assess the influence of segment
interactions on the mechanical response
of an incompressible polymer network at equi-biaxial tension.

\subsection{Equi-biaxial tension}

Equi-biaxial tension of an incompressible material is
described by the formulas
\[
x_{1}=kX_{1},
\qquad
x_{2}=kX_{2},
\qquad
x_{3}=k^{-2} X_{3},
\]
where $k$ stands for elongation ratio.
The right Cauchy--Green tensor reads
\[
{\bf C}=k^{2}({\bf e}_{1}{\bf e}_{1}
+{\bf e}_{2}{\bf e}_{2})
+k^{-4} {\bf e}_{3}{\bf e}_{3},
\]
and its eigenvalues are given by
\begin{equation}
\lambda_{1}=\lambda_{2}=k^{2},
\qquad
\lambda_{3}=k^{-4}.
\end{equation}
As equi-biaxial tests on an incompressible layer
are conventionally performed in the tensile mode,
we confine ourselves to the case $k>1$.
Substituting expressions (73) into Eqs. (65) and
using Eq. (64), we find that
\begin{equation}
\bar{W} = \frac{k_{B}TM}{2}\biggl [
\Bigl (1+\frac{\varepsilon}{3} \sqrt{\frac{6}{\pi}}\Bigr )
\Bigl (2 k^{2}+\frac{1}{k^{4}}-3\Bigr )
-\varepsilon\sqrt{\frac{6}{\pi}}
\Bigl ( \frac{1}{k^{2}} +\frac{k^{4}}{\sqrt{k^{6}-1}}
\arcsin \frac{\sqrt{k^{6}-1}}{k^{3}}-2\Bigr )\biggr ].
\end{equation}
According to Eqs. (66), the Cauchy tensile stress $\Sigma$
is determined by the formula
\[
\Sigma=\frac{1}{2}\lambda_{1}
\frac{\partial \bar{W}}{\partial \lambda}
-\lambda_{3}\frac{\partial \bar{W}}{\partial \lambda_{3}}
=\frac{k}{2}\frac{d \bar{W}}{dk}.
\]
It follows from this relation that the engineering tensile
stress $\sigma=\Sigma/k$ reads
\begin{equation}
\sigma=\frac{1}{2}\frac{d \bar{W}}{dk}.
\end{equation}
Substitution of expression (74) into Eq. (75) results in
\begin{equation}
\sigma = k_{B}TM \biggl \{
\Bigl (1+\frac{\varepsilon}{3} \sqrt{\frac{6}{\pi}}\Bigr )
\Bigl (k-\frac{1}{k^{5}}\Bigr )
+\frac{\varepsilon}{4}\sqrt{\frac{6}{\pi}}
\Bigl [ \frac{2}{k^{3}}-\frac{k^{3}}{k^{6}-1}
\Bigl (\frac{k^{6}-4}{\sqrt{k^{6}-1}}
\arcsin \frac{\sqrt{k^{6}-1}}{k^{3}}+3\Bigr )\Bigr ]
\biggr \}.
\end{equation}
It follows from Eq. (76) that $\lim_{k\to 1} \sigma(k)=0$,
which means that the initial state is stress-free.

The dependence of the dimensionless tensile stress
$\sigma_{0}$ on elongation ratio $k$ is depicted in
Figure 3 for $\varepsilon=0$ (a network of Gaussian chains)
and $\varepsilon=0.5$
(a network of flexible chains with excluded-volume
interactions).
This figure shows that segment interactions induce
an increase in the tensile stress at all elongation
ratios $k$.
The difference between tensile stresses monotonically
grows with $k$.
At $k=6.0$ (which is in the range of deformations reached
in experiments on elastomers), the engineering stress in
a network of chains with segment interactions exceeds
that in a network of Gaussian chains by 28\%.

The results of numerical simulation at uniaxial tension
of an incompressible medium are also presented in Figure 3
for comparison.
According to this figure, at relatively large
elongation ratios ($k>3.0$), the difference between
the tensile stresses (corresponding to these two
deformation modes) in a Gaussian network disappears,
whereas an appropriate difference in a network of
flexible chains with excluded-volume interactions
remains quite pronounced.
Although under both deformation programs, excluded-volume
interactions induce an increase in stiffness of
the network, the effect of segment interactions on
the stress--strain relation is stronger at equi-biaxial
tension than at uniaxial tension.

\section{Concluding remarks}

The concept of non-entropic elasticity is proposed
for the analysis of the mechanical response of a network
of flexible chains with excluded-volume interactions.
Unlike the classical theory of entropic elasticity,
where the free energy of a chain is entirely characterized
by the distribution function of end-to-end vectors,
we associate the free energy of a chain with the average
value of its Hamiltonian.
The average free energy of a chain and the strain
energy density of a network are calculated explicitly
(under the assumption that the strength of excluded-volume
interactions is small compared with thermal energy).
Constitutive equations are developed for a network
of macromolecules with intra-chain interactions under
an arbitrary three-dimensional deformation.
These relations are simplified for uniaxial tension
(compression) and equi-biaxial tension of an
incompressible medium at finite strains.
An explicit expression is derived for the elastic
modulus of a network of chains with weak segment
interactions.
The effect of intra-chain interactions on the
stress--strain diagram is analyzed numerically.
It is demonstrated that excluded-volume interactions
result in an increase in the tensile stress at
both deformation modes under consideration.
This growth is substantial (about 30\% at
elongations typical for experiments on rubbers),
which means that the account for excluded-volume
interactions is quite important for applications.

\newpage
\section*{List of figures}
\parindent 0 mm

{\bf Figure 1:}
The dimensionless reduced stress $\sigma_{\ast}$
versus elongation ratio $k$ at uniaxial tension.
\vspace*{2 mm}

{\bf Figure 2:}
The dimensionless reduced stress $\sigma_{\ast}$
versus elongation ratio $k$ at uniaxial compression.
\vspace*{2 mm}

{\bf Figure 3:}
The dimensionless tensile stress $\sigma_{0}$
versus elongation ratio $k$ at uniaxial (filled circles)
and equi-biaxial (solid lines) deformations.
\vspace*{100 mm}

\setlength{\unitlength}{0.75 mm}
\begin{figure}[tbh]
\begin{center}

\end{center}
\vspace*{10 mm}

\caption{}
\end{figure}

\setlength{\unitlength}{0.75 mm}
\begin{figure}[tbh]
\begin{center}
%
\end{center}
\vspace*{10 mm}

\caption{}
\end{figure}

\setlength{\unitlength}{0.75 mm}
\begin{figure}[tbh]
\begin{center}
%
\end{center}
\vspace*{10 mm}

\caption{}
\end{figure}

\end{document}